\documentclass{appolb}
\usepackage[utf8x]{inputenc}

\usepackage{amsmath,amsfonts,amsthm,amssymb}
\usepackage[dvips]{graphics,graphicx}
\usepackage[usenames,dvipsnames]{color}
\definecolor{darkblue}{RGB}{0,0,196}
\definecolor{darkgreen}{RGB}{0,120,0}
\usepackage[colorlinks=true,linktocpage=true,linkcolor=darkblue,citecolor=red,urlcolor=darkblue]{hyperref}
\usepackage{cancel}
\usepackage{bbold}
\usepackage{multirow}
\usepackage{longtable}
\usepackage{color}
\usepackage{xcolor}
\usepackage[normalem]{ulem}
\usepackage{hyperref}
\usepackage{bigints}
\usepackage{xparse}
\usepackage{physics}
\usepackage{verbatim}
\usepackage{minibox}
\usepackage{comment}
\usepackage{appendix}
\usepackage{slashed}
\usepackage{todonotes}

\usepackage{marginnote}
\usepackage{graphicx}
\usepackage[nice]{nicefrac}
\usepackage{amsmath}
\usepackage{hepunits}
\def\be{\begin{equation}}
\def\ee{\end{equation}}
\newcommand{\bel}[1]{\begin{eqnarray}\label{#1}}
\newcommand{\eel}{\end{eqnarray}}
\def\barr{\begin{array}}
\def\earr{\end{array}}
\def\beq{\begin{eqnarray}}
\def\eeq{\end{eqnarray}}
\def\bfig{\begin{figure}}
\def\efig{\end{figure}}

\newcommand{\nn}{\nonumber}
\def\LR{\left(} % round
\def\RR{\right)}
\def\LS{\left[} % square
\def\RS{\right]}

\newcommand{\sh}[1]{\sinh#1}
\newcommand{\ch}[1]{\cosh#1}

\newcommand{\xv}{{\boldsymbol x}}

\def\S0iU{{\Sigma}^{0i}}

\def\n0{n_{0}}
\def\e0{\varepsilon_{0}}
\def\P0{P_{0}}

\def\be{\begin{equation}}
	\def\ee{\end{equation}}

\def\barr{\begin{array}}
	\def\earr{\end{array}}
\def\beq{\begin{eqnarray}}
	\def\eeq{\end{eqnarray}}
\def\bfig{\begin{figure}}
	\def\efig{\end{figure}}
\newcommand{\bea}{\begin{eqnarray}}
	\newcommand{\eea}{\end{eqnarray}}

\def\LR{\left(} % round
\def\RR{\right)}
\def\LS{\left[} % square
\def\RS{\right]}

\newcommand{\refeq}[1]{Eq.~(\ref{#1})}

\def\pv{{\boldsymbol p}}

\def\be{\begin{equation}}
\def\ee{\end{equation}}
\def\ba{\begin{eqnarray}}
\def\ea{\end{eqnarray}}   

\def\LR{\left(} % round
\def\RR{\right)}
\def\LS{\left[} % square
\def\RS{\right]}

\def\pv{{\boldsymbol p}}

\def\n0{n_{(0)}}
\def\e0{\varepsilon_{(0)}}
\def\P0{P_{(0)}}

\usepackage{stackengine,scalerel}
\newcommand\hstar[1]{\ThisStyle{\ensurestackMath{%
  \setbox0=\hbox{$\SavedStyle#1$}%
  \stackengine{0pt}{\copy0}{\kern.2\ht0\smash{\SavedStyle\star}}{O}{c}{F}{T}{S}}}}

\definecolor {darkgreen}{rgb}{0.2,0.7,0.2}

\def\LR{\left(} % round
\def\RR{\right)}
\def\LS{\left[} % square
\def\RS{\right]}
\def\nn{\nonumber}

\newcommand{\refeqb}[1]{(\ref{#1})} % blank

\def\ev{{\boldsymbol e}}
\def\pv{{\boldsymbol p}}
\def\xv{{\boldsymbol x}}
\def\rv{{\boldsymbol r}}

\begin{document}

\title{Deuteron production in a combined thermal and coalescence framework for heavy-ion collisions  in
the few-GeV energy regime}

\author{Wojciech Florkowski, Piotr Salabura, Nikodem Witkowski
%\email{wojciech.florkowski@uj.edu.pl}
\address{Jagiellonian University, PL-30-348 Krak\'ow, Poland}\\
\medskip 
Radoslaw Ryblewski
%\email{radoslaw.ryblewski@ifj.edu.pl}
\address{Institute  of  Nuclear  Physics  Polish  Academy  of  Sciences,  PL-31-342  Krak\'ow,  Poland}\\
}

\date{Received: date / Accepted: date}

\maketitle

\begin{abstract}
A recently formulated thermal model for hadron production in heavy-ion collisions in the few-GeV energy regime is combined with the idea that some part of protons and neutrons present in the original thermal system forms deuterons via the coalescence mechanism. Using realistic parametrizations of the freeze-out conditions, which reproduce well the spectra of protons and pions, we make predictions for deuteron transverse-momentum and rapidity spectra. The best agreement with the experimentally known deuteron yield is obtained if the freeze-out temperature is relatively high and, accordingly, the system size at freeze-out is rather small. In addition, the standard Gaussian distribution of the relative distance between nucleons is replaced by the distribution resulting from their independent and approximately uniform production inside the initial thermal system.
\end{abstract}
%%%%%%%%%%%%%%%%%%%%%%%%%%%%%%%%%%%%%%%%%
\section{Introduction}
%%%%%%%%%%%%%%%%%%%%%%%%%%%%%%%%%%%%%%%%%
%
Recently, a statistical hadronization model of hadron production in heavy-ion collisions in the few-GeV energy regime has been formulated that very well describes the transverse-momentum and rapidity spectra of protons and pions (both positive and negative)~\cite{Harabasz:2020sei,Harabasz:2022rdt}. The main ingredient of this new approach is an assumption about \emph{spherical}~\cite{Harabasz:2020sei} or slightly \emph{spheroidal}~\cite{Harabasz:2022rdt} expansion of the produced matter. This assumption should be contrasted with the common use of boost-invariant blast-wave models at ultrarelativistic energies~\cite{Schnedermann:1993ws}, which becomes inappropriate for the description of heavy-ion collisions at the beam energies of a few GeV per nucleon pair. 

The idea of spherical expansion of the produced matter can be traced back to the original formulation of the blast-wave model by Siemens and Rasmussen~\cite{Siemens:1978pb} (see also Ref.~\cite{Florkowski:2010zz}). Clearly, at lower energies, the assumption about spherical expansion is more realistic than a boost-invariant scenario that implies a~constant rapidity distribution. However, our results have shown that  spheroidal expansion (with a slightly larger longitudinal flow compared to the transverse one) significantly improves description of the experimental data~\cite{Harabasz:2022rdt}.

\begin{table*}[t]
    \centering
    \begingroup
    \setlength{\tabcolsep}{13pt} % Default value: 6pt
    \renewcommand{\arraystretch}{1.2} % Default value: 1
    \begin{tabular}{|c|c|c|c|}
    \hline
    \hline
    Parameter & Spherical& A & B \\
    \hline
    $T$ (MeV) & $49.6$ & $49.6$ & $70.3$ \\
    $\mu_B$ (MeV) & $776$  & $776$ & $876$ \\
    $\mu_{I_3}$ (MeV) & $-14.1$  & $-14.1$ & $-21.5$ \\
    \hline
    $R$ (fm) & $ 16.02 $  & $ 15.7 $ & $6.06$ \\
    $H$ (MeV) & $8.0$  & $10.0$ & $22.5$ \\
    $\delta$ & $0$  & $0.2$ & $0.4$ \\
    \hline
    $v_R = \hbox{tanh}(HR) $ & 0.57 & 0.66 & 0.60 \\
    $\gamma_R = \ch(HR)$ & 1.22 & 1.33 & 1.25 \\
    \hline
    \hline
    \end{tabular}
    \endgroup
    \caption{Upper part: thermodynamic parameters obtained from different fitting strategies to ratios of particle multiplicities measured in Au-Au collisions at the beam energy $\sqrt{s_{\rm_{NN}}}=2.4$ GeV~\cite{Harabasz:2020sei,Harabasz:2022rdt,Motornenko:2021nds}. Middle part: the system's radius $R$, the Hubble-like expansion parameter $H$, and momentum-space longitudinal eccentricity $\delta$ obtained from the fits to the experimentally measured proton and pion spectra~\cite{Harabasz:2022rdt}. Lower part: the radial flow and Lorentz gamma factor at the system's boundary ($r=R$).}
    \label{tab:params}
\end{table*}

In this work, we continue the analysis of the data collected by the HADES
Collaboration for Au-Au collisions at the beam energy $\sqrt{s_{\rm NN}}=$~2.4 GeV and the centrality class of 10\%~\cite{Szala:2019,Szala2019a,HADES:2020ver}. The fits to the particle abundances done by our and other groups~\cite{Motornenko:2021nds} suggest two different sets of possible freeze-out thermodynamic parameters. The main difference between them resides in two different values of the freeze-out temperature: $T=$~49.6~MeV vs. $T=$~70.3~MeV. In the following, we refer to these two cases as the low- and high-temperature ones. They are labeled by $``A"$ and $``B"$, respectively. In these two cases, we assume a spheroidal expansion of matter at freeze-out. In addition, we consider a spherically symmetric case with the (low) freeze-out temperature $T=$~49.6~MeV, which was introduced for the first time in Ref.~\cite{Harabasz:2020sei}. Hence, we consider here altogether three different freeze-out scenarios. Their thermodynamic parameters (temperature, baryon, and isospin chemical potentials) are listed in the upper part of Table~\ref{tab:params}. The other thermodynamic parameters that control strangeness production are not displayed here as they are irrelevant to the present analysis.

In the considered collisions of gold nuclei, one detects about 28.7 nuclei of $^2$H, 8.7 nuclei of $^3$H, and 4.6 nuclei of $^3$He~\cite{Szala:2019,Szala2019a}.Hence, about  46.5 protons are found in the bound states. The remaining average number of directly produced protons is 77.6. In the framework defined in Refs.~\cite{Harabasz:2020sei,Harabasz:2022rdt} we assume that all protons eventually detected in bound states and those measured as free particles originally constitute a thermal system. Thus, while comparing the predictions of our model with the data, the final results for the proton spectra are rescaled by the factor 77.6/(77.6+46.5). This implicitly assumes the existence of a certain physical mechanism that combines the remaining protons with neutrons into the light nuclei mentioned above. In the present work, we assume that this process can be interpreted as coalescence. 

The standard expressions of the coalescence framework (see, for example, Ref.~\cite{Mrowczynski:2016xqm}) are used to predict the deuteron spectra in the three different cases introduced above. Taking into account the experimentally known deuteron yield, we find that the coalescence model is more consistent with the high-temperature scenario. In this case, the system's size $R$ is rather small, of about 6~fm, hence the probability of forming deuteron by a proton-neutron pair is relatively large (compared to the low-temperature case where $R \approx 16$~fm). A novel feature of our approach is that we do not use Gaussian \textit{ad hoc} parametrizations of the thermal source but rather we consider initial thermal particle production as independent pair production within a sphere of the radius $R$ (which is consistent with thermal model assumptions adopted in Refs.~\cite{Harabasz:2020sei,Harabasz:2022rdt}).

Our present work does not include contributions from decaying resonances. We have checked using THERMINATOR~\cite{Kisiel:2005hn,Chojnacki:2011hb} that for the low-temperature case, they are negligible, while for the high-temperature case, only the Delta resonance plays a noticeable, yet very small, role. In fact, the results presented in this work favor the high-temperature freeze-out, which indicates that the presence of the Delta resonance in the thermal approach should be examined in more detail, with an emphasis on differences caused by the use of different forms of the Delta density of states~\cite{Lo:2017ldt}. This issue is planned for our future investigations.

The topic of producing deuteron and other light nuclei has recently attracted a lot of interest (see, for example, Refs.~\cite{Sombun:2018yqh,Blum:2019suo,Mrowczynski:2019yrr,Bellini:2020cbj,Kachelriess:2020amp,Kozhevnikova:2020bdb,Zhao:2020irc,Hillmann:2021zgj,ALICE:2021mfm,Sochorova:2021lal,Zhao:2021dka,Sharma:2022tih}). In relativistic heavy-ion collisions, the production of such systems is well reproduced by both the thermal approach and the coalescence model. These two approaches are usually treated as exclusive alternatives (for a short and critical review of this issue see Ref.~\cite{Mrowczynski:2020ugu}). In this work we present a scenario for lower energies where the thermal initial production of particles is combined with a subsequent coalescence mechanism in a consistent way, showing that the two pictures may coexist and supplement each other.

The paper has the following structure: In Sec.~\ref{sec:coal} we introduce the coalescence model and discuss the deuteron formation rate. In Sec.~\ref{sec:fm} the freeze-out model is defined. The spherical and spheroidal forms of expansion are discussed in Sec.~\ref{sec:sph} and Sec.~\ref{sec:spd}, respectively. In these two sections, we introduce a compact description of the distribution functions with given symmetries, which turns out to be very useful for the implementation of the coalescence model. We conclude in Sec.~\ref{sec:summary}. Throughout the paper, we use natural units: $c=\hbar=k_B=1$. The metric tensor has the signature $(+1,-1,-1,-1)$.

%
%%%%%%%%%%%%%%%%%%%%%%%%%%%%%%%%%%%%%%%%%
\section{Coalescence model}
\label{sec:coal}
%%%%%%%%%%%%%%%%%%%%%%%%%%%%%%%%%%%%%%%%%
%
%%%%%%%%%%%%%%%%%%%%%%%%%%%%%%%%%%%%%%%%%
\subsection{Basic concept}
%%%%%%%%%%%%%%%%%%%%%%%%%%%%%%%%%%%%%%%%%
%
The basic idea of the coalescence model for deuteron production is that the deuteron spectrum is obtained as the product of proton and neutron spectra taken at half of the deuteron momentum~\cite{Butler:1963pp,Schwarzschild:1963zz}. To be more specific, we define the proton and neutron three-momentum distributions by the functions
\begin{equation}
F_p({\pv}) = \frac{dN_p}{d^3p}, \quad F_n({\pv}) = \frac{dN_n}{d^3p}, 
\label{eq:FpFn}
\end{equation}
and the deuteron distribution as the product
\begin{equation}
\frac{dN_d}{d^3p_d} = A_{\rm FR} \,
F_p \left(\frac{\pv_d}{2} \right) F_n \left(\frac{\pv_d}{2} \right),
\label{eq:coal}
\end{equation}
where $A_{\rm FR}$ is the deuteron formation rate coefficient discussed in more detail below and the subscripts $d, p, n$ refer to deuterons, protons, and neutrons, respectively

Equation (\ref{eq:coal}) assumes that $\pv_d = 2 \pv$, hence, additivity of three-momenta of nucleons forming a deuteron. Classically, this leads to a small violation of the energy conservation: due to the finite deuteron binding energy $m_d = 1876$~MeV, while $m_p+m_n = (938+940)$~MeV = 1878~MeV. This and other related problems connected with the conservation laws in the coalescence model are usually circumvented by reference to the quantum nature of the real coalescence process, which introduces natural uncertainty of the energies and momenta of interacting particles~\cite{Mrowczynski:2016xqm}. In any case, small differences between $m_d$ and $m_p+m_n$, as well as between $m_p$ and $m_n$ do not affect our present analysis. Therefore, in the following, we use the approximation $m_p \approx m_n \approx m$, where $m$ is the mean nucleon mass, and $m_d \approx 2m$. Hence, while switching in \refeq{eq:coal} from the variable $\pv_d$ to the deuteron transverse momentum $p_{\perp d}$ and rapidity $y_d$ we use simple rules
\begin{equation}
p_\perp = \frac{p_{\perp d}}{2} ,  
\quad  
y = y_d .
\label{eq:pperpdyd}
\end{equation}

As in both theory and experiment one usually deals with invariant momentum distributions, $E \,dN/(d^3p)$, rather than with the forms (\ref{eq:FpFn}), it is convenient to recall that for cylindrically symmetric  (with respect the beam axis $z$) systems studied in this work we have
\begin{equation}
\frac{dN}{d^3p} = \frac{dN}{2\pi\, E\, dy\, p_\perp dp_\perp} =
\frac{dN}{2\pi E\, dy\, m_\perp dm_\perp},
\end{equation}
where $E$ is the on-mass-shell energy of a particle $E=\sqrt{m^2+\pv^2}$ and $m_\perp$ is its transverse mass $m_\perp=\sqrt{m^2+p_\perp^2}$. Therefore, from \refeq{eq:coal} we obtain
\begin{eqnarray}
\frac{dN_d}{E_d\, dy\, m_{\perp d} dm_{\perp d}} = \frac{A_{\rm FR}}{2\pi}
\frac{dN_p}{E\, dy\, m_{\perp} dm_{\perp}}
\frac{dN_n}{E\, dy\, m_{\perp} dm_{\perp}} .
\label{eq:coalymt}
\end{eqnarray}
Note that since the momenta and energies of protons and neutrons are equal, the proton and neutron distributions differ only due to different thermodynamic parameters used in equilibrium distribution functions. Note also that within our approximations $y_d=y_p=y_n=y$ and at zero rapidity \refeq{eq:coalymt} reduces to the formula
\begin{eqnarray}
\left. \frac{dN_d}{dy\, m_{\perp d}^2 dm_{\perp d}} 
\right|_{y=0} &=& \frac{A_{\rm FR}}{2\pi}
\left.\frac{dN_p}{dy\, m_{\perp}^2 dm_{\perp}}
\right|_{y=0} 
\left.\frac{dN_n}{dy\, m_{\perp}^2 dm_{\perp}} 
\right|_{y=0} .
\label{eq:coaly0}
\end{eqnarray}
In our previous works, we have constructed two models that reproduce well the spectra of protons and neutrons~\cite{Harabasz:2020sei,Harabasz:2022rdt}. Here we can use the same models to predict, based on \refeq{eq:coaly0}, the deuteron spectrum. For finite values of rapidity \refeq{eq:coalymt} takes the form
\begin{eqnarray}
\frac{dN_d}{dy\, m_{\perp d}^2 dm_{\perp d}} 
 &=& \frac{A_{\rm FR}}{2\pi \ch y}
 \frac{dN_p}{dy\, m_{\perp }^2 dm_{\perp }} 
\frac{dN_n}{dy\, m_{\perp }^2 dm_{\perp }} .
\label{eq:coaly}
\end{eqnarray}
By integration of this equation over the deuteron transverse mass, we obtain the deuteron rapidity distribution.

%%%%%%%%%%%%%%%%%%%%%%%%%%%%%%%%%%%%%%%%%
\subsection{Deuteron formation rate}
%%%%%%%%%%%%%%%%%%%%%%%%%%%%%%%%%%%%%%%%%
%

A popular form of the coefficient $A_{\rm FR}$ used in the literature is~\cite{Mrowczynski:2016xqm}
\begin{eqnarray}
A_{\rm FR} = \frac{3}{4} (2\pi)^3 \int d^3r \,D(r) \, \left| \phi_d(r)\right|^2.
\label{eq:A}
\end{eqnarray}
Here the function $D(r)$ is the normalized to unity distribution of the relative spacetime positions of the neutron and proton pairs at freeze-out, while $\phi_d(r)$ is the deuteron wave function of relative motion. Many works on the coalescence model assume a Gaussian profile \cite{Mrowczynski:1992gc,
Mrowczynski:1987oid} 
\begin{equation}
D(r) = \left(4 \pi R_{\rm kin}^2 \right)^{-3/2}
\exp\left(- \frac{r^2}{4 R_{\rm kin}^2} \right),
\label{eq:DrG}
\end{equation}
where $R_{\rm kin}$ is the radius of the system at freeze-out. This formula may be, however, regarded as inconsistent with the physical assumptions used in our model, where original particles are produced independently inside a~sphere of radius $R$ at a fixed laboratory time $t$~\cite{Harabasz:2020sei,Harabasz:2022rdt}. Expression (\ref{eq:DrG}) gives the root-mean-squared value $r_{\rm rms} = \sqrt{6} R \approx 2.45 R$, which implies deuteron production far away from the original thermal system and its long formation time. Thus, as an alternative to the Gaussian distribution (\ref{eq:DrG}), we use the distribution of a relative distance for particles produced independently in a~sphere of radius $R$ (for more details see Appendix \ref{sec:app})
\begin{equation}
D(r) = \frac{3}{4\pi R^3} \left(1 - \frac{3 r}{4 R} + \frac{r^3}{16 R^3} \right)
\theta_H(2 R - r).
\label{eq:DrS}
\end{equation}
Here $\theta_H(x)$ is the Heaviside step function. It naturally restricts the relative distance between two members of a pair to $r \leq 2 R$. Below, we occasionally refer to \refeq{eq:DrS} as to the sharp cutoff distribution.

Since our system strongly expands, the particle densities become larger with growing distance $r$ from the center. This is controlled by the Lorentz gamma factor $\gamma(r) =\cosh(H r)$. Implementation of relativistic corrections to \refeq{eq:DrS} is described in Appendix \ref{sec:app}. We have found that for our values of $H$ and $R$ the implementation of such corrections leads to small effects, on the order of a few percent, and can be neglected. 

\begin{table*}[t]
    \centering
    \begingroup
    \setlength{\tabcolsep}{13pt} % Default value: 6pt
    \renewcommand{\arraystretch}{1.2} % Default value: 1
    \begin{tabular}{|c|c|c|c|}
    \hline
    \hline
    Formation rate & Spherical& A & B \\
    \hline
    $A_{\rm GG}$ (MeV$^3$) & $7 \,565$ & $8 \,028$ & $120 \,509$ \\
    $A_{\rm SG}$ (MeV$^3$) & $64 \,239$  & $67 \,860$ & $693 \,463$ \\
    $A_{\rm SH}$ (MeV$^3$) & $69 \,661$  & $73 \,735$ & $942 \,476$ \\
    \hline
    \hline 
    \end{tabular}
    \endgroup
    \caption{Values of the formation rate parameter $A_{\rm FR}$ for different choices of the functions $D(r)$ and $\phi_d(r)$: $A_{\rm GG}$ is obtained with the two Gaussian profiles, Eqs.~(\ref{eq:DrG}) and (\ref{eq:phidG}),  and $R_{\rm kin} = R$;  $A_{\rm SG}$ follows from Eqs.~(\ref{eq:DrS}) and (\ref{eq:phidG}); finally, $A_{\rm SH}$ is calculated with Eqs.~(\ref{eq:DrS}) and (\ref{eq:phidH}).
    }
    \label{tab:As}
\end{table*}
For the deuteron wave function, we use two options: i) the Gaussian approximation
\begin{equation}
\left| \phi_d(r)\right|^2 = \left(4 \pi R_d^2 \right)^{-3/2}
\exp\left(- \frac{r^2}{4 R_d^2} \right),
\label{eq:phidG}
\end{equation}
where $R_d = 2.13~$fm is the deuteron radius, and ii) the Hulthen wave function defined by the expression~\cite{Bellini:2020cbj}
\begin{equation}
\phi_d(r) = \sqrt{\frac{\alpha \beta (\alpha+\beta)}{2\pi (\alpha-\beta)^2}}
\frac{\exp\left(-\alpha r\right) - \exp\left(-\beta r\right)}{r},
\label{eq:phidH}
\end{equation}
where $\alpha = 0.2~$fm$^{-1}$ and $\beta = 1.56~$fm$^{-1}$.~\footnote{We use here traditional notation, $\beta$ appearing in \refeq{eq:phidH} should not be confused with inverse temperature.} Both $D(r)$ and $\left| \phi_d(r)\right|^2$ are normalized to unity. 

\begin{figure}[t] 
\centering
\includegraphics[width=0.7\textwidth]{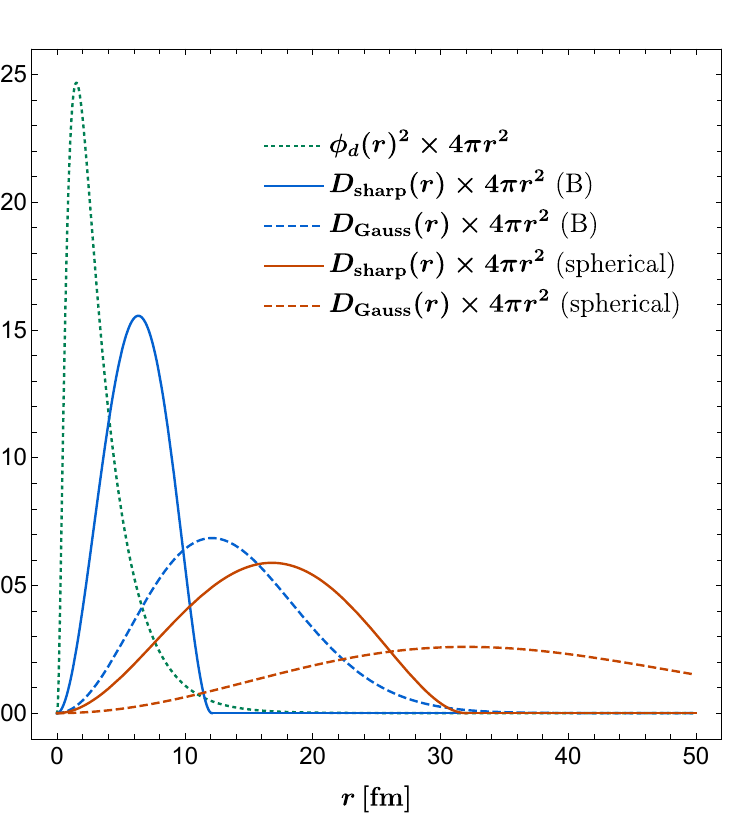} 
\caption{The square of the Hulthen wave function and different versions of the nucleon pair distribution function $D(r)$ multiplied by the factor $4\pi r^2$.  }
\label{fig:figurephi}
\end{figure}

The values of the formation rate coefficient for different freeze-out scenarios are shown in Table~\ref{tab:As}. The value obtained with the two Gaussian profiles (Eqs.~(\ref{eq:DrG}) and (\ref{eq:phidG})) is denoted by $A_{\rm GG}$, the sharp cutoff distribution~(\ref{eq:DrS}) combined with the Gaussian or Hulthen wave function gives $A_{\rm SG}$ or $A_{\rm SH}$, respectively. We observe that the values of $A_{\rm FR}$ do not significantly differ for the spherical and spheroidal $A$ cases -- they are both low-temperature scenarios with large freeze-out radii. However, an increase in the magnitude of $A_{\rm FR}$ is clearly seen if we switch from the Gaussian to the sharp cutoff distribution of pairs. An additional increase of the magnitude of $A_{\rm FR}$ is seen if we switch to the spheroidal $B$ scenario. In this case, the freeze-out radius is relatively small ($\sim 6$~fm) and the overlap of the pair distribution with the deuteron wave function becomes the largest. Due to uncertainties connected with the estimate of $A_{\rm FR}$, we present below our results for the three different options defined in Table~\ref{tab:As}. The Hulthen wave function squared and different forms of the functions $D(r)$ are shown in Fig.~\ref{fig:figurephi}.

%%%%%%%%%%%%%%%%%%%%%%%%%%%%%%%%%%%%%%%%%
\section{Freeze-out models}
\label{sec:fm}
%%%%%%%%%%%%%%%%%%%%%%%%%%%%%%%%%%%%%%%%%
%
The freeze-out models specify the hydrodynamic conditions for particle production at the latest stages of the system's spacetime evolution. In the single-freeze-out scenario~\cite{Broniowski:2001we,Broniowski:2001uk} adopted here, the freeze-out stage is defined by a set of thermodynamic variables such as temperature $T$ and baryon chemical potential $\mu_B$, and the shape of the freeze-out hypersurface $\Sigma$. In addition, one defines the form of the hydrodynamic
flow $u^\mu(x)$ on $\Sigma$.

%
%%%%%%%%%%%%%%%%%%%%%%%%%%%%%%%%%%%%%%%%%
\subsection{Cooper-Frye formula}
%%%%%%%%%%%%%%%%%%%%%%%%%%%%%%%%%%%%%%%%%
%

The standard starting point for quantitative calculations is the Cooper-Frye formula \cite{Cooper:1974mv} that describes the invariant momentum spectrum of particles
\begin{equation}
E \frac{dN}{d^3p} = \int d^3\Sigma_\mu(x) \, p^\mu f(x,p). 
\label{cf}
\end{equation}
Here $f(x,p)$ is the phase-space distribution function of particles, and $p^\mu = \LR E, \pv \RR$ is their four-momentum with the mass-shell energy $E = \sqrt{m^2 + \pv^2}$.

The infinitesimal element of a three-dimensional freeze-out hypersurface from which particles are emitted $d^3\Sigma_\mu(x)$ may be obtained from the formula (see, for example, Ref.~\cite{Misner:1973prb})
\begin{equation}
d^3\Sigma_\mu = -\epsilon_{\mu \alpha \beta \gamma}
\frac{\partial x^\alpha}{\partial a }
\frac{\partial x^\beta }{ \partial b }
\frac{\partial x^\gamma }{ \partial c }\, da\, db\, dc,
\label{d3sigma}
\end{equation}
where $\epsilon_{\mu \alpha \beta \gamma}$ is the Levi-Civita tensor with the convention $\epsilon_{0123}=-1$ and $a, b, c$ are the three independent coordinates introduced to parametrize the hypersurface. This allows us to construct a six-dimensional, Lorentz invariant density of the produced particles
\begin{equation}
   d^{\,6}N    =   \frac{d^3p}{E} \,\,d^3\Sigma \cdot p \,\, f(x,p).
\label{cf2}
\end{equation}
The independent variables in such a general parametrization would be three components of three-momentum and the variables $a$, $b$, and $c$.

%
%%%%%%%%%%%%%%%%%%%%%%%%%%%%%%%%%%%%%%%%%
\subsection{Local equilibrium distributions}
%%%%%%%%%%%%%%%%%%%%%%%%%%%%%%%%%%%%%%%%%
%
%
In local equilibrium, the distribution function $f(x,p)$ has the following general form
\begin{equation}
f(x,p)= \frac{g_s}{(2\pi)^3}\LS \Upsilon^{-1} \exp\LR\frac{p \cdot u}{T}\RR-\chi \RS^{-1},
\label{feq}
\end{equation}
where $T$ is the freeze-out temperature, $\chi=-1$ ($\chi=+1$) for Fermi-Dirac (Bose-Einstein) statistics, and $g_s = 2s +1$ is the degeneracy factor connected with spin. In this work we use distributions of protons and neutrons only (as input for the deuteron coalescence formula), hence we take $\chi=-1$ and $s=1/2$. Since we use the same mass for protons and neutrons, their equilibrium distributions differ only by the values of thermodynamic potentials that determine the fugacity factor
\begin{eqnarray}
\Upsilon_p =  \exp \left( \frac{\mu_B + \frac{1}{2} \mu_{I_3}}{T}\right), \nn \\
\Upsilon_n =  \exp \left( \frac{\mu_B - \frac{1}{2} \mu_{I_3}}{T}\right).
\label{upsiNeq}
\end{eqnarray}
Here $\mu_B$ and $\mu_{I_3}$ are baryon and isospin chemical potentials, respectively. The values of $T$, $\mu_B$, and $\mu_{I_3}$ are given in Table~\ref{tab:params} for different expansions scenarios.

%
%%%%%%%%%%%%%%%%%%%%%%%%%%%%%%%%%%%%%%%%%
\section{Spherical expansion}
\label{sec:sph}
%%%%%%%%%%%%%%%%%%%%%%%%%%%%%%%%%%%%%%%%%
%
%%%%%%%%%%%%%%%%%%%%%%%%%%%%%%%%%%%%%%%%%
\subsection{Symmetry implementation}
%%%%%%%%%%%%%%%%%%%%%%%%%%%%%%%%%%%%%%%%%
%

For spherically symmetric freeze-outs, it is convenient to introduce the following parametrization of the space-time points on the freeze-out hypersurface 
\begin{equation}
x^\mu = (t,\xv) = \LR t(\zeta), r(\zeta) \ev_r \RR, 
\label{xsphere} 
\end{equation}
where $\ev_r = \LR  \cos\phi\sin\theta,  \sin\phi \sin\theta,  \cos\theta \RR$. The freeze-out hypersurface is completely defined if a curve, i.e., the mapping $\zeta \longrightarrow \left(t(\zeta),r(\zeta)\right)$ in the $t-r$ space is given. This curve defines the (freeze-out) times $t$ when the hadrons in the shells of radius $r$ stop interaction. The range of $\zeta$ may be always restricted to the interval: $0 \leq \zeta \leq 1$.  The three coordinates: $\phi \in [0,2 \pi), \theta \in [0,\pi]$, and $\zeta \in [0,1]$ play the role of the variables $a, b, c$ appearing in \refeq{d3sigma}. Hence, the element of the spherically symmetric hypersurface has the form 
\begin{eqnarray}
\! d^3\Sigma_\mu \!&=& \!
\LR r^\prime(\zeta), t^\prime(\zeta)\ev_r  \RR
r^2(\zeta) \sin\theta \,d\theta\,d\phi\,d\zeta,
\label{d3sigmasphere}
\end{eqnarray}
where the prime denotes the derivatives taken with respect to $\zeta$. 

Besides the spherically symmetric hypersurface, we introduce the spherically symmetric hydrodynamic flow
\begin{eqnarray}
u^\mu &=& \gamma(\zeta)\LR 1,
v(\zeta)\ev_r  \RR
\label{usphere}
\end{eqnarray}
where $\gamma(\zeta)$ is the Lorentz factor, which due to the normalization condition $u\cdot u=1$ is given by the formula $\gamma(\zeta) = (1-v^2(\zeta))^{-1/2}$. In a similar way, the four-momentum of a hadron is parameterized as
\begin{equation}
p^\mu = \LR E, p\, \ev_p \RR,
\label{psphere}
\end{equation}
with $\ev_p = \LR  \cos\phi_p \sin\theta_p, \sin\phi_p \sin\theta_p,  \cos\theta_p \RR$. Thus, we find that 
\begin{equation}
 u \cdot p=\gamma(\zeta) \LR E - p\, v(\zeta) \kappa \RR
\label{pusphere}
\end{equation}
and
\begin{equation}
d^3\Sigma \cdot p = \LR E_p r^\prime(\zeta) 
- p\, t^\prime(\zeta)  \kappa \RR r^2(\zeta) \sin\theta
\,d\theta \, d\phi \, d\zeta,
\label{pdSigmashere}
\end{equation}
where $\kappa\equiv \ev_p \cdot \ev_r= \cos\theta \cos\theta_p + \sin\theta \sin\theta_p \cos(\phi-\phi_p)$.

Our previous analyses of the HADES data showed that a good description of the experimental results can be obtained with the freeze-out hypersurface defined by the condition $t=$~const (in this case the variable $\zeta$ may be identified with the distance $r$) and with the hydrodynamic radial flow of the form
\begin{equation}
v(r) = \tanh(H r),
\label{eq:hubble}
\end{equation}
where $H$ plays a role of the Hubble constant~\cite{Chojnacki:2004ec}. In this case, the Cooper-Frye formula for fermions takes the form
\begin{eqnarray}
\frac{dN}{dy m_\perp^2 dm_\perp} &=& \frac{g_s \ch y }{(2\pi)^2} \int\limits_0^R dr \, r^2 \int\limits_0^\pi d\theta \sin\theta \int\limits_0^{2\pi} d\phi \nn \\
&& \hspace{-1cm} \times \left[\Upsilon^{-1}\exp\left( \frac{\gamma(r) (E\!-\!p \, v(r) \kappa)}{T} \right) + 1 \right]^{-1}.
\label{eq:sph0}
\end{eqnarray}
Due to spherical symmetry, the integral on the RHS of \refeq{eq:sph0} is independent of the angles $\theta_p$ and $\phi_p$, hence we may set $\theta_p = \phi_p =0$ ($\kappa = \cos\theta$) and write
\begin{eqnarray}
\frac{dN}{dy \, m_\perp^2 dm_\perp} &=& \ch y \,\, S(p)
\label{eq:sph1}
\end{eqnarray}
where 
\begin{eqnarray}
S(p) &=& \frac{g_s }{2\pi} \int\limits_0^R dr \, r^2 \int\limits_0^\pi d\theta \sin\theta  \label{eq:sph2} \\
&& \hspace{-1cm} \times \left[\Upsilon^{-1}\exp\left( \frac{ E \ch(H r)\!-\!p \, \sh(H r)\cos\theta}{T} \right) + 1 \right]^{-1}. \nn
\end{eqnarray}
The values of $R$ and $H$ are given in Table~\ref{tab:params}. For the spherical expansion, the radial velocity at the system's boundary is $v_R = \tanh(H R) \approx 0.57$, and the Lorentz gamma factor $\gamma_R = \cosh(H R) \approx 1.22$. In the case of the Boltzmann statistics, the integral over the angle $\theta$ in Eq.~(\ref{eq:sph2}) is analytic and we find
\begin{eqnarray}
S(p) &=& \frac{g_s \Upsilon}{\pi} \int\limits_0^R dr \, r^2 
 \exp\left(-\frac{ E \ch(H r)}{T} \right) 
 \frac{\sinh\left(  a \right)}{a} ,
\label{eq:sph2B}
\end{eqnarray}
where $a = (p/T) \sh(H r)$. 

%%%%%%%%%%%%%%%%%%%%%%%%%%%%%%%%%%%%%%%%%
\subsection{Rapidity and transverse-mass distributions}
%%%%%%%%%%%%%%%%%%%%%%%%%%%%%%%%%%%%%%%%%

%%%%%%%%%%%%%%%%%%%%%%%%%%%%%%%%%%%%%%%%%
\subsubsection{Protons}
In the spherical case, our results for protons and neutrons depend only on the magnitude of their three-momentum
\begin{equation}
p = \sqrt{p_x^2 + p_y^2 + p_z^2} = \sqrt{p_\perp^2 + m_\perp^2 \sh^2 y}.  
\end{equation}
Hence, the transverse-momentum distribution of protons or neutrons at zero rapidity is directly given by the function $S(p_\perp)$, namely
\begin{equation}
\left. \frac{dN_{p,n}}{dy\, m_\perp^2 dm_\perp}  \right|_{y=0} = S_{p,n}(p_\perp).
\label{eq:Sperp}
\end{equation}
On the other hand, the rapidity distribution is given by the integral
\begin{equation}
\frac{dN}{dy} = \ch y \int\limits_m^\infty S\left(\sqrt{p_\perp^2 + m_\perp^2 \sh^2 y} \right)\, m_\perp^2 dm_\perp.
\end{equation}

\begin{figure}[t]\includegraphics[width=0.49\textwidth]{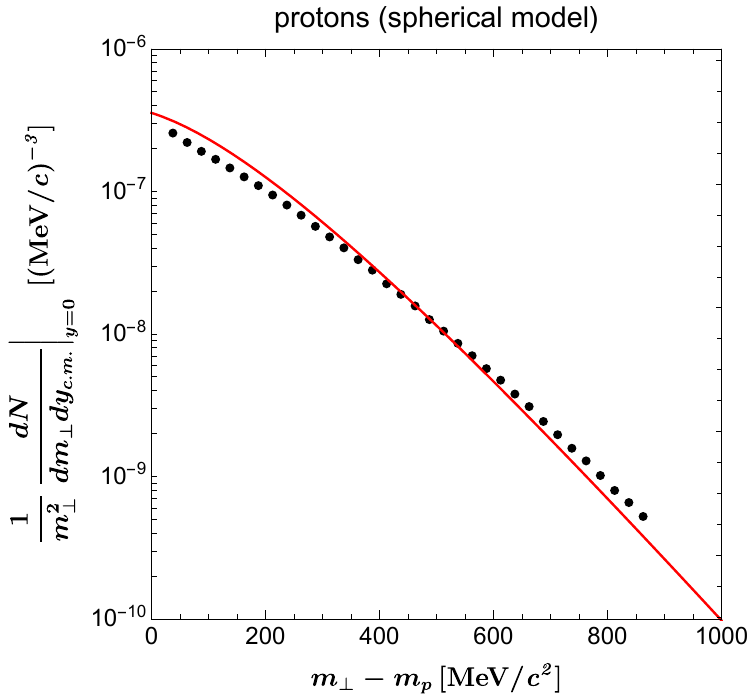} 
\includegraphics[width=0.49\textwidth]{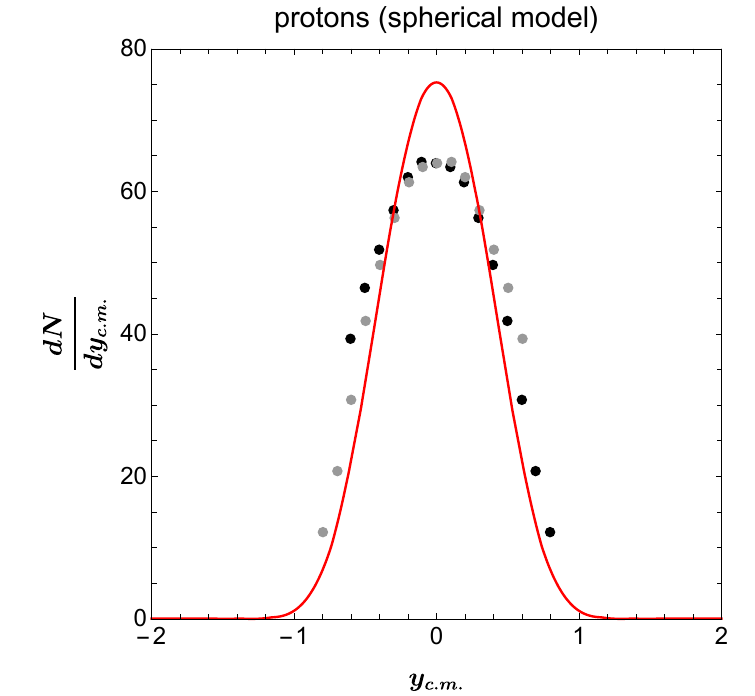} % 
\caption{Transverse-momentum ({\bf left}) and rapidity ({\bf right}) spectra of protons obtained in the spherical model (solid red lines) compared with the HADES data.}
\label{fig:figure1}
\end{figure}

Our results for the transverse-momentum and rapidity distributions of protons in the spherical model are shown in Fig.~\ref{fig:figure1}. We reproduce the results obtained first in Ref.~\cite{Harabasz:2022rdt}. The model parameters have been fitted to the transverse-momentum spectra of hadrons, hence, we observe that the proton transverse-momentum distribution is reproduced better (left panel of Fig.~\ref{fig:figure1}) than the rapidity distribution (right panel of Fig.~\ref{fig:figure1}). This deficiency is overcome in the spheroidal model discussed below. The total yield of protons $N_{p}$ in the spherical model is $72.0$, while $(dN_{p}/dy)_{y=0}\approx 75.3$. The experimental result is $N_{p} = 77.6$, hence differs by less than 10\%.

%%%%%%%%%%%%%%%%%%%%%%%%%%%%%%%%%%%%%%%%%%%%%%%%%%%%%%%%%
\subsubsection{Deuterons}
Having the proton model spectra reproduced, we can turn to the analysis of the deuteron production. In this case, we use \refeq{eq:pperpdyd} and \refeq{eq:Sperp}, and rewrite \refeq{eq:coaly0} in a compact form as
\begin{eqnarray}
\left. \frac{dN_d}{dy\, m_{\perp d}^2 dm_{\perp d}} 
\right|_{y=0} &=& \frac{A_{\rm FR}}{2\pi}
S_p\left(\frac{p_{\perp d}}{2} \right)
S_n\left(\frac{p_{\perp d}}{2} \right), \nn \\
\label{eq:coaly0sph}
\end{eqnarray}
where we can use the substitution $p_{\perp d} = \sqrt{m_{\perp d}^2 - m_d^2}$. For finite values of rapidity, we use
\begin{eqnarray}
\frac{dN_d}{dy\, m_{\perp d}^2 dm_{\perp d}} 
 &=& \frac{A_{\rm FR}}{2\pi \ch y}
 S_p\left( \frac{
 \sqrt{m_{\perp d}^2 \ch^2 y - m_d^2}
 }{2} \right)
 \nn \\
&& \hspace{-1.5cm}\times
S_n\left( \frac{
\sqrt{m_{\perp d}^2 \ch^2 y - m_d^2}
}{2}  \right).
\label{eq:coaly}
\end{eqnarray}
\begin{figure}[t]\includegraphics[width=0.49\textwidth]{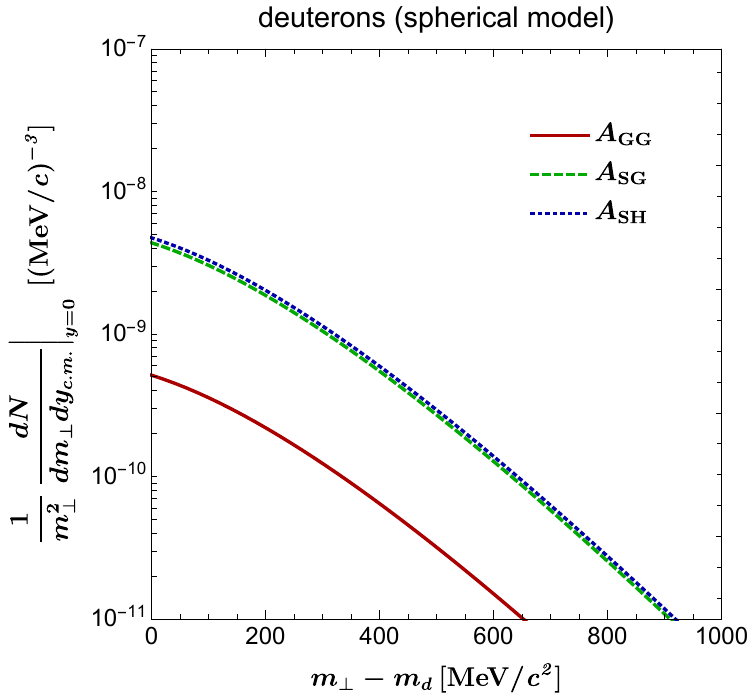} \includegraphics[width=0.49\textwidth]{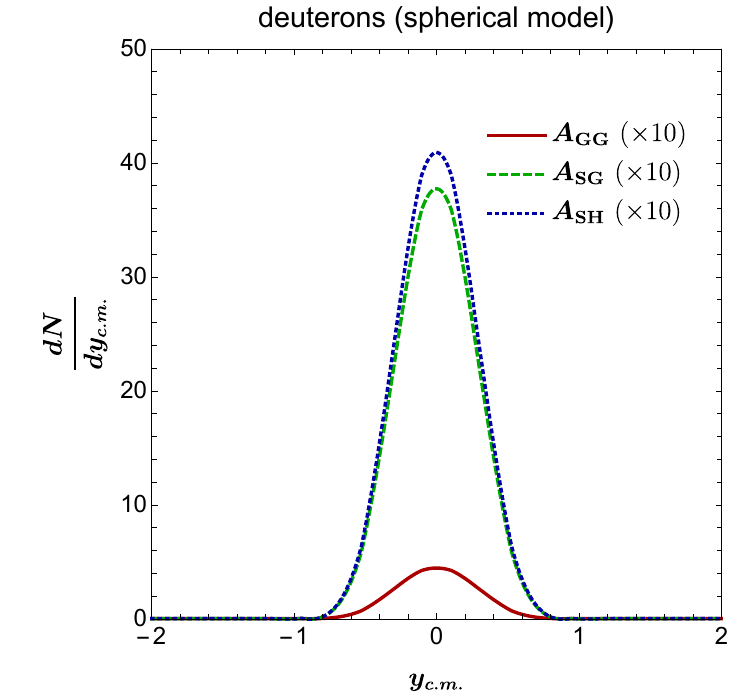}
\caption{Predictions of the spherical model for the deuteron production. {\bf Left:} model transverse-momentum spectra obtained for three different values of the formation rate coefficient $A_{\rm FR}$ (as given in Table~\ref{tab:As}). {\bf Right:} model rapidity distributions, again for three different choices of $A_{\rm FR}$. The model rapidity distributions in the right panel are multiplied by the factor of 10.}
\label{fig:figure2}
\end{figure}

The predictions of the spherical model for the deuteron spectra are shown in Fig.~\ref{fig:figure2}. Although the use of the sharp cutoff distribution for pairs increases the value of $A_{\rm FR}$ by approximately a factor of 10, compared to the Gaussian distribution, there is still a factor of 10 missing to reproduce the correct normalization of the spectra. We obtain: 
$N_{d} \approx  0.31$,  $N_{d} \approx  2.66$, and $N_{d} \approx  2.88$  for $A_{\rm GG}$, $A_{\rm SG}$, and $A_{\rm SH}$, respectively. The corresponding rapidity densities are: 
$(dN_{d}/dy)_{y=0} \approx 0.44$,
$(dN_{d}/dy)_{y=0} \approx 3.77$, and
$(dN_{d}/dy)_{y=0} \approx 4.09$.
The measured deuteron yield is 28.7.

%%%%%%%%%%%%%%%%%%%%%%%%%%%%%%%%%%%%%%%%%
\section{Spheroidal expansion}
\label{sec:spd}
%%%%%%%%%%%%%%%%%%%%%%%%%%%%%%%%%%%%%%%%%
%
%%%%%%%%%%%%%%%%%%%%%%%%%%%%%%%%%%%%%%%%%
\subsection{Symmetry implementation}
%%%%%%%%%%%%%%%%%%%%%%%%%%%%%%%%%%%%%%%%%
%
%
For spheroidally symmetric freeze-outs with respect to the beam axis, it is convenient to introduce the following parametrization of the space-time points on the freeze-out hypersurface 
% %
% \begin{equation}
% x^\mu = \LR t, r \sqrt{1-\epsilon}\cos\phi\sin\theta, r \sqrt{1-\epsilon}\sin\phi \sin\theta, r \sqrt{1+\epsilon}\cos\theta \RR,
% \label{xspheroid} 
% \end{equation} 
% %
\begin{equation}
x^\mu = \LR t, r \sqrt{1-\epsilon}\,\ev_{r\perp}, r \sqrt{1+\epsilon}\,
\cos\theta  \RR.
\label{xspheroid} 
\end{equation}
Here the parameter $\epsilon$ controls deformation from a spherical shape, while $\ev_{r\perp} = \LR  \cos\phi\sin\theta,  \sin\phi \sin\theta\RR$. For $\epsilon~>~0$ the hypersurface is stretched in the (beam) $z$-direction.  The resulting infinitesimal element of the spheroidally symmetric hypersurface has the form 
\begin{eqnarray}
\! d^3\Sigma_\mu \!&=& (1-\epsilon)\!\LR 
r^\prime  \sqrt{1+\epsilon}, 
t^\prime \frac{\sqrt{1+\epsilon}}{\sqrt{1-\epsilon}}\,\ev_{r\perp},
t^\prime \,\cos\theta \RR
r^2  \sin\theta \,d\theta\,d\phi\,d\zeta,\nn\\
\label{d3sigmaspheroid}
\end{eqnarray}
where the prime denotes the derivatives taken with respect to $\zeta$. Besides the spheroidally symmetric hypersurface, we introduce the spheroidally symmetric hydrodynamic flow
\begin{eqnarray}
u^\mu &=& \gamma(\zeta,\theta)\LR 1,
v(\zeta) \sqrt{1-\delta}\,\ev_{r\perp}, 
v(\zeta) \sqrt{1+\delta}\,\cos\theta \RR, 
\label{uspheroid}
\end{eqnarray}
where $\gamma(\zeta,\theta)$ is the Lorentz factor, which due to the normalization condition $u\cdot u=1$ is given by the formula 
\begin{equation}
\gamma(\zeta,\theta) = \LS 1-(1+\delta \cos(2 \theta))v^2(\zeta)\RS^{-1/2}.
\end{equation}
Thus, we find that 
\begin{equation}
 u \cdot p=\gamma(\zeta,\theta) \LR E\!-\!p\, v(\zeta) \kappa (\delta) \RR
\label{puspheroid}
\end{equation}
and
\begin{equation}
d^3\Sigma \cdot p = \sqrt{1-\epsilon}\LR E \,r^\prime \sqrt{1-\epsilon^2} 
- p \,t^\prime   \kappa(-\epsilon) \RR r^2 \sin\theta
\,d\theta \, d\phi \, d\zeta,
\label{Sigmapspheroid}
\end{equation}
where $\kappa (\xi)=   \sqrt{1+\xi}\cos\theta \cos\theta_p + \sqrt{1-\xi}\sin\theta \sin\theta_p \cos(\phi-\phi_p)$.
It is trivial to check that the formula \refeqb{Sigmapspheroid} agrees with \refeqb{pdSigmashere} if $\epsilon=0$, as well as \refeq{puspheroid} agrees with \refeq{pusphere} for $\delta=0$.

Our earlier analysis of the spectra showed that a very good description of the data can be obtained by assuming $t^\prime=0$ (that allows for using $\zeta = r$) and $\epsilon=0$, however with $\delta \neq 0$. Then, we have, as in the spherical case
\begin{equation}
d^3\Sigma \cdot p =  E \, r^2  dr \sin\theta
\,d\theta \, d\phi .
\label{Sigmapspheroidtp0}
\end{equation}
The Cooper-Frye formula for fermions takes the form
\begin{eqnarray}
\frac{dN}{dy \, m_\perp^2 dm_\perp} &=& \ch y \,\, {\tilde S}(p,\theta_p)
\label{eq:spd1}
\end{eqnarray}
where 
\begin{eqnarray}
{\tilde S}(p,\theta_p)\!&=\!&\frac{g_s }{(2\pi)^2} \int\limits_0^R dr \, r^2 \int\limits_0^\pi d\theta \sin\theta  \int\limits_0^{2\pi} d\phi  \left[\Upsilon^{-1} \exp\left( \frac{ u\cdot p }{T} \right) + 1 \right]^{-1}. \label{eq:tS} 
\end{eqnarray}
With $u\cdot p$ given by \refeq{puspheroid}, where due to the spheroidal symmetry we can set $\phi_p = 0$. In the case of the Boltzmann statistics, the integral over the angle $\phi$ is analytic and yields the modified Bessel function of the first kind $I_0$. The approximate expression for the function ${\tilde S}(p,\theta_p)$ in this case is given by a two-dimensional integral
\begin{eqnarray}
{\tilde S}_B(p,\theta_p)\!&=\!& \frac{g_s \Upsilon}{2\pi}  \int\limits_0^R dr \, r^2 \!\! \int\limits_0^\pi d\theta \sin\theta  
\exp\left( -\frac{\gamma (E\!-\!\sqrt{1+\delta} \, p v \cos\theta \cos\theta_p)}{T} \right) \nn \\ 
&& \times I_0\left(\frac{\gamma \sqrt{1-\delta} \, p v \sin\theta \sin\theta_p}{T} \right). \label{eq:tSB} 
\end{eqnarray}

%%%%%%%%%%%%%%%%%%%%%%%%%%%%%%%%%%%%%%%%%
\subsection{Rapidity and transverse-mass distributions}
%%%%%%%%%%%%%%%%%%%%%%%%%%%%%%%%%%%%%%%%%
%
%%%%%%%%%%%%%%%%%%%%%%%%%%%%%%%%%%%%%%%%%
\subsubsection{Protons}

Finally, by changing variables from $p$ and $\theta_p$ to rapidity and transverse mass, we may write
\begin{eqnarray}
\frac{dN}{dy \, m_\perp^2 dm_\perp} &=& \ch y \,\, 
{\tilde S}\left[ \sqrt{m_\perp^2 \cosh^2 y -m^2}, \theta_y(m_\perp,y) \right],
\label{eq:spd2}
\end{eqnarray}
where
\begin{equation}
\theta_y(m_\perp, y)  = \arccos 
\frac{m_\perp \sh y}{\sqrt{m_\perp^2 \cosh^2 y -m^2}}.
\label{eq:thetay}
\end{equation}
We note that within our approximations the angle (\ref{eq:thetay}) is the same for nucleons and deuterons. At zero rapidity we obtain as the special case
\begin{eqnarray}
\left. \frac{dN}{dy \, m_\perp^2 dm_\perp} \right|_{y=0 }&=&  
{\tilde S}\left(p_\perp, \frac{\pi}{2} \right).
\label{eq:spd3}
\end{eqnarray}
\begin{figure}[ht]\includegraphics[width=0.495\textwidth]{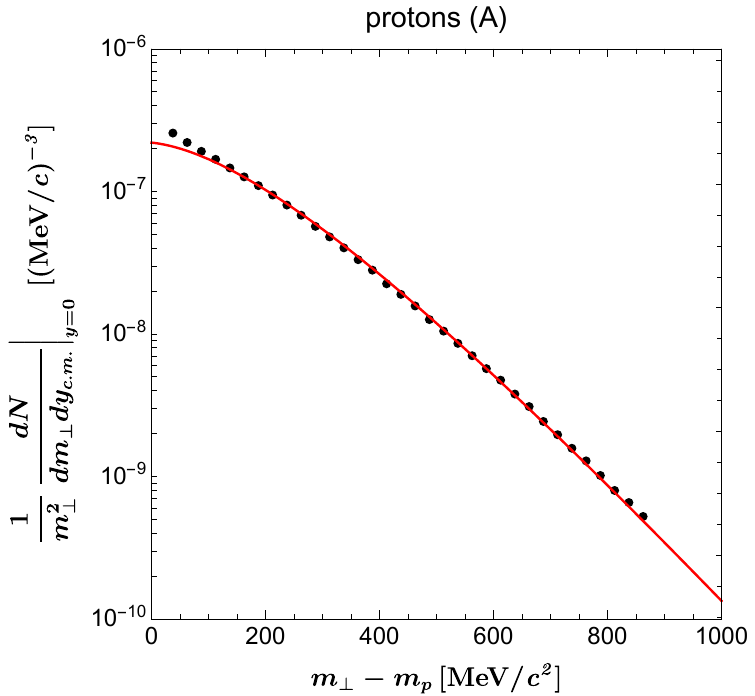} 
\includegraphics[width=0.495\textwidth]{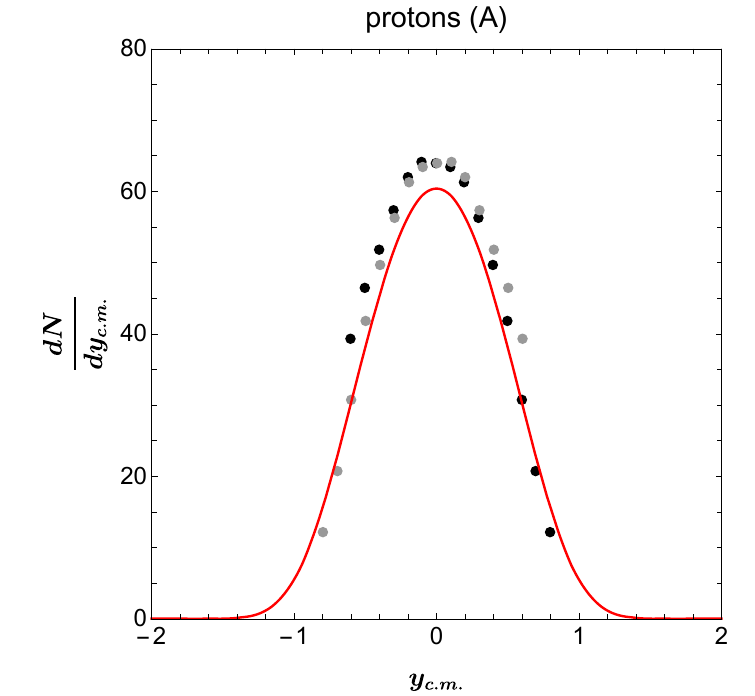} % 
\caption{Transverse-momentum ({\bf left}) and rapidity ({\bf right}) spectra of protons obtained in the spherical model version $A$ (solid red lines) compared with the HADES data.}
\label{fig:figure3}
\end{figure}
\begin{figure}[ht]\includegraphics[width=0.495\textwidth]{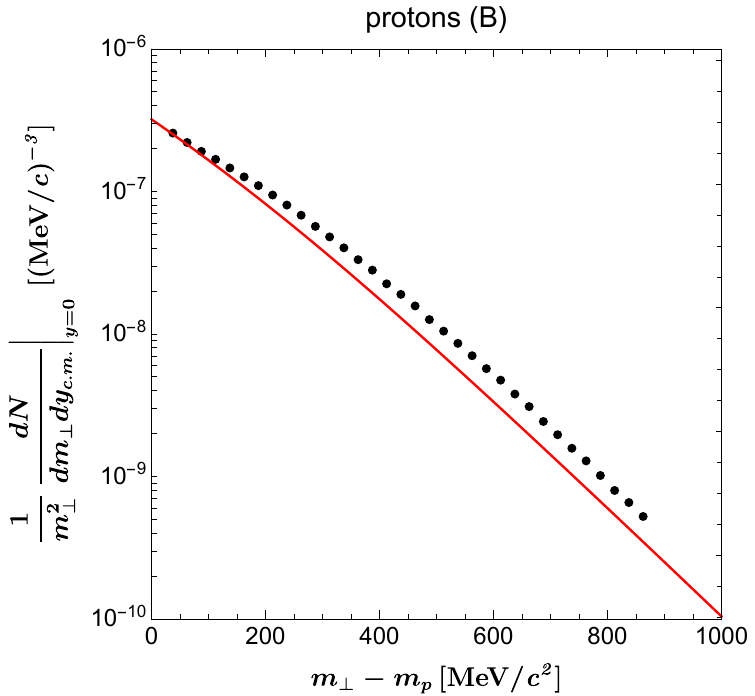} 
\includegraphics[width=0.495\textwidth]{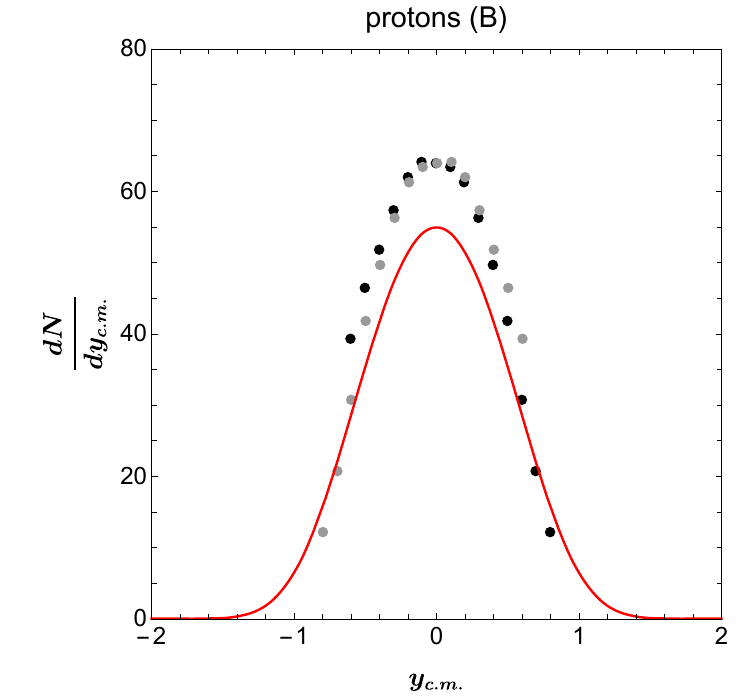} % 
\caption{Same as Fig.~\ref{fig:figure3} but for the spheroidal model version $B$.  }
\label{fig:figure4}
\end{figure}
 
Our results for the transverse-momentum and rapidity distributions of protons in the spheroidal model are shown in Figs.~\ref{fig:figure3} and~\ref{fig:figure4} for versions $A$ and $B$, respectively. We reproduce here again the results of Ref.~\cite{Harabasz:2022rdt}. We emphasize that the spheroidal model is able to consistently describe both the transverse-momentum and rapidity spectra. The model results for the proton yield and rapidity density at $y=0$ are: $N_{p} \approx 73.78$, $(dN_{p}/dy)_{y=0}\approx 60.35$ for the version $A$, and $N_{p} \approx 69.35$, $(dN_{p}/dy)_{y=0}\approx 54.90$ for the version $B$.

%
%%%%%%%%%%%%%%%%%%%%%%%%%%%%%%%%%%%%%%%%%
\subsubsection{Deuterons}

Having checked that we can reproduce the proton spectra, we can make predictions for the deuterons. In this case we use \refeq{eq:coaly} with the nucleon spectrum defined by Eq.~(\ref{eq:spd2}). 

\begin{figure}[t]\includegraphics[width=0.495\textwidth]{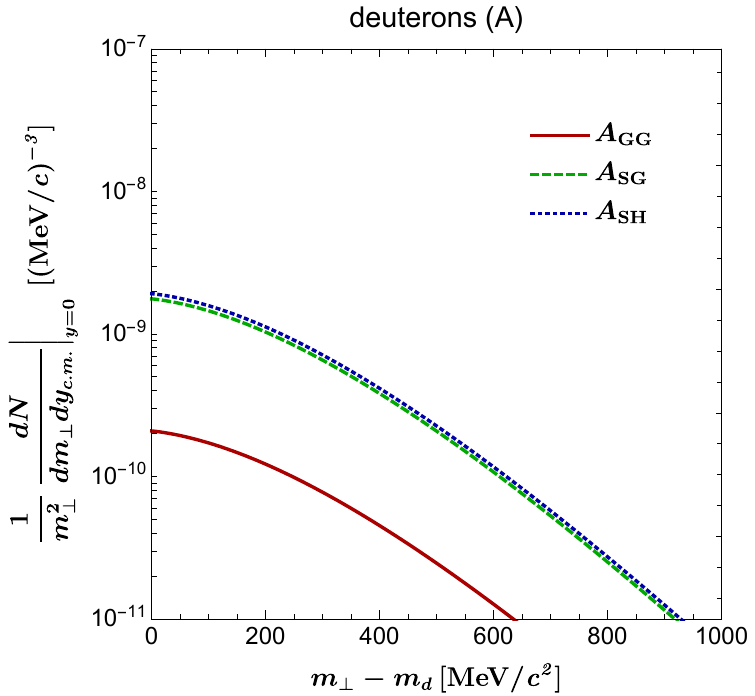} 
\includegraphics[width=0.495\textwidth]{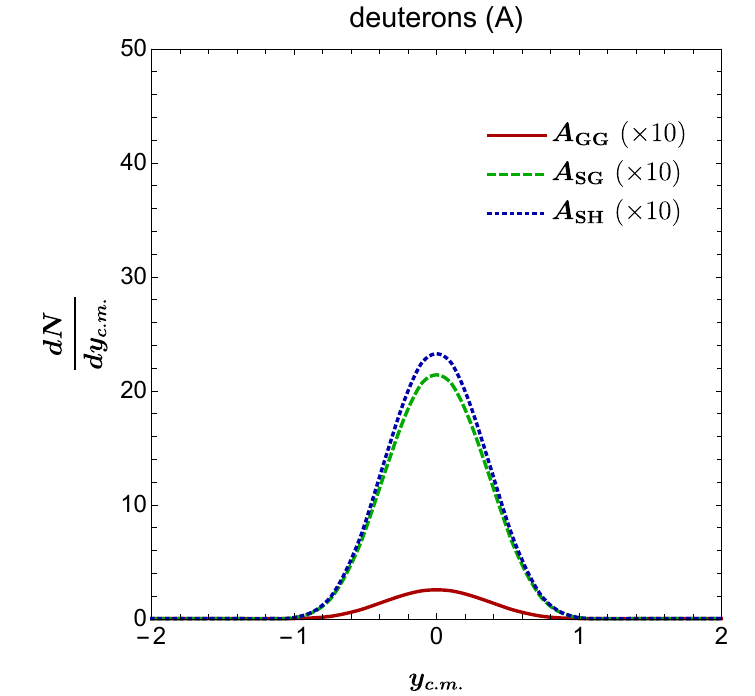} % 
\caption{Predictions for the deuteron spectra in the spheroidal model version $A$. The coding of the lines is the same as in Fig.~\ref{fig:figure2}. The model rapidity distribution is multiplied by a factor of 10.}
\label{fig:figure5}
\end{figure}
\begin{figure}[t]
\includegraphics[width=0.495\textwidth]{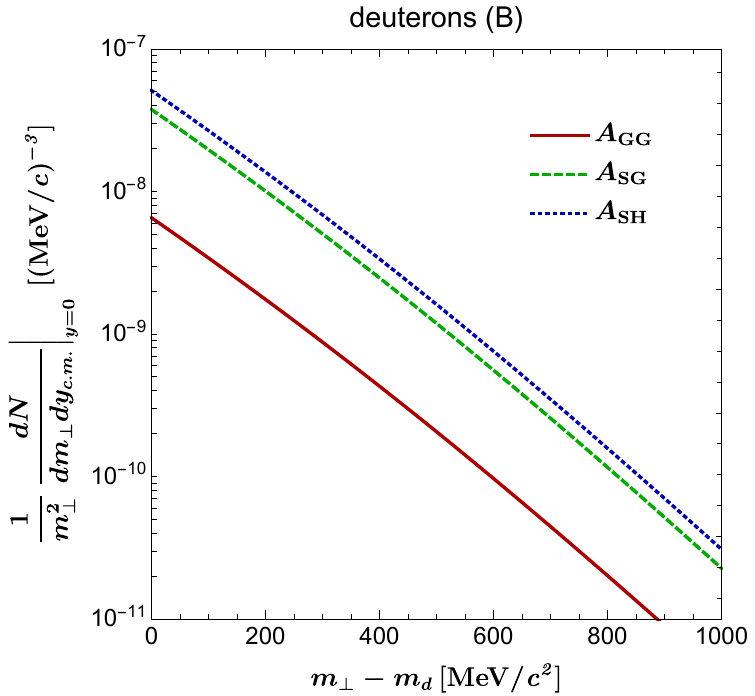} 
\includegraphics[width=0.495\textwidth]{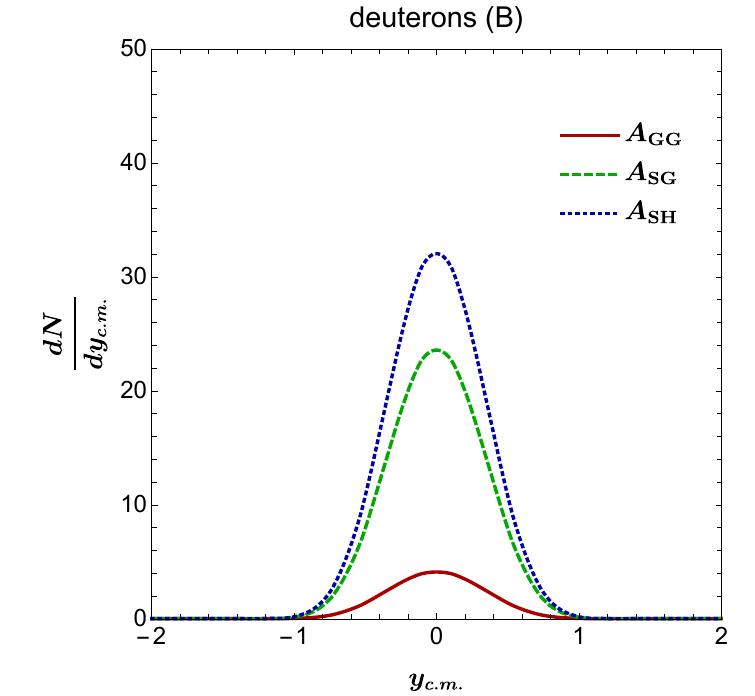} 
\caption{Predictions for the deuteron spectra in the spheroidal model version $B$. The coding of the lines is the same as in Fig.~\ref{fig:figure2}.}
\label{fig:figure6}
\end{figure}

In the spheroidal model $A$ we obtain the following deuteron yields: $N_{d}~\approx~0.22$, $N_{d} \approx 1.86$, and $N_{d} \approx 2.02$ for three different choices of the formation rate coefficient. These values are two (for $A_{\rm GG}$) or one (for $A_{\rm SG}$ and $A_{\rm SH}$) order of magnitude smaller than the measured value $N_{d} = 28.7$. The corresponding rapidity densities are:
$(dN_{d}/dy)_{y=0} \approx 0.25$,
$(dN_{d}/dy)_{y=0} \approx 2.14$, and
$(dN_{d}/dy)_{y=0} \approx 2.32$.

The results for the spheroidal model $B$ are: 
$N_{d} \approx 3.46$, $N_{d} \approx 19.89$, 
$N_{d} \approx 27.04$ for the yields, and
$(dN_{d}/dy)_{y=0} \approx 4.09$,
$(dN_{d}/dy)_{y=0} \approx 23.56$, and
$(dN_{d}/dy)_{y=0} \approx 32.02$ for the rapidity density at $y=0$. Larger values for the version $B$ are a direct consequence of the smaller system's size at freeze-out. The result obtained with $A_{\rm SH}$ is very close to the experimental result $N_{d} = 28.7 \pm 0.8$.

%%%%%%%%%%%%%%%%%%%%%%%%%%%%%%%%%%%%%%%
\section{Summary and conclusions}
\label{sec:summary}
%%%%%%%%%%%%%%%%%%%%%%%%%%%%%%%%%%%%%%%

In this work, we have extended our statistical hadronization model for heavy-ion collisions in the few-GeV energy regime~\cite{Harabasz:2020sei,Harabasz:2022rdt} to include description of deuteron production via a coalescence mechanism. Already in Refs.~\cite{Harabasz:2020sei,Harabasz:2022rdt} we assumed that nucleons detected in the bound states are originally present in a thermal system. However, no explicit calculation was given as to how they become bound states. The framework offered in this work fills this gap. 

We find that the slope of the transverse-momentum spectra of deuterons follows naturally from the main coalescence ansatz that the deuteron spectrum is the product of nucleon spectra taken at half of the deuteron three-momentum. However, the normalization of the deuteron spectrum depends very strongly on the value of the so-called formation rate coefficient. The latter may vary from 7 565 MeV$^3$ (for the spherical model at low temperature with the Gaussian distribution of pairs in space and the Gaussian approximation for the deuteron wave function) to 942 476 MeV$^3$ (for the spheroidal model at high temperature, more realistic distributions of pairs, and the Hulthen deuteron wave function).

Both, a higher freeze-out temperature (a smaller system's size) and a non-Gaussian distribution of the distance between the original pairs forming the deuteron increase the probability that a nucleon pair forms a deuteron. Each of these effects increases the formation rate by a factor of 10.  

Taking into account the measured yield of deuterons, our present work favors the freeze-out scenario at a higher freeze-out temperature combined with a spheroidal expansion. This case may be further examined by a study of other interesting aspects such as the contribution from the Delta resonance, Lambda spin polarization (as in Ref.~\cite{Florkowski:2019voj}), and the production of other light nuclei.
 
\section*{Acknowledgments}
N.W. would like to thank Prof.~Tetyana Galatyuk for the financial support that made it possible to start this project at GSI. The authors also thank Małgorzata Gumberidze and Szymon Harabasz for fruitful discussions of the model assumptions. This work was supported in part by the Polish National Science Centre Grants No. 2022/47/B/ST2/01372 (W.F.) and No. 2018/30/E/ST2/00432 (R.R.).

%%%%%%%%%%%%%%%%%
%
\begin{appendix}
\section{Distribution of the relative distance between pairs created independently and uniformly in a sphere}
\label{sec:app}
The two-particle distribution of pairs produced independently and uniformly within a sphere with radius $R$ is given by the product of the two-step functions
\begin{equation}
f(\rv_1, \rv_2) = \left( \frac{3}{4 \pi R^3} \right)^2
\theta_H(R-|\rv_1|) \theta_H(R-|\rv_2|).
\label{eq:fr1r2}
\end{equation}
The function $f(\rv_1, \rv_2)$ satisfies the normalization condition
\begin{equation}
\int f(\rv_1, \rv_2) d^3r_1 d^3r_2 = 1.
\end{equation}
By introducing the mean and relative position vectors: $\rv = (\rv_1+\rv_2)/2$, $\delta \rv = \rv_1 - \rv_2$, we rewrite the normalization condition as
\begin{equation}
\int  \left( \frac{3}{4 \pi R^3} \right)^2
\theta_H\left(R-\left|\rv +\frac{\delta \rv}{2}\right| \right) 
\theta_H\left(R-\left|\rv -\frac{\delta \rv}{2}\right| \right) 
d^3r  \, d^3 \delta r.
\end{equation}
From this expression, we identify the relative distance distribution of the pairs as

\begin{equation}
D(\delta r) = \int  \left( \frac{3}{4 \pi R^3} \right)^2
\theta_H\left(R-\left|\rv +\frac{\delta \rv}{2}\right| \right) 
\theta_H\left(R-\left|\rv -\frac{\delta \rv}{2}\right| \right) 
d^3r  .
\label{eq:reld}
\end{equation}
Due to the rotational invariance, the function $D(\delta r)$ depends only on the length of $\delta\rv$. The calculation of the integral (\ref{eq:reld}) can be reduced to a geometric problem described, for example, in Sec. 3 of~\cite{Fetter:1973}, with the final result given by \refeq{eq:DrS}.

Due to relativistic expansion, the density of particles grows with the distance from the center. This growth is controlled by the Lorentz gamma factor $\cosh(H r)$. To include this effect in our considerations, we may generalize \refeq{eq:fr1r2} to the form
\begin{eqnarray}
f_r(\rv_1, \rv_2) &=& N^2_r(R,H) 
\cosh\left(H |\rv_1|\right) \theta_H(R-|\rv_1|) \nn \\
&& \times
\cosh\left(H |\rv_2|\right) \theta_H(R-|\rv_2|).
\label{eq:relfr1r2}
\end{eqnarray}
The normalization constant $N_r$ is determined by the condition
\begin{eqnarray}
1 =  N_r(R,H) \int_0^R \cosh\left(H r\right) 4 \pi r^2\, dr
\end{eqnarray}
and is given by the expression
\begin{equation}
N^{-1}_r(R,H) = \frac{4\pi}{H^3} \left[\left(2+H^2 R^2 \right)\sinh(HR) -
2 H R \cosh(H R) \right].
\end{equation}
For small values of $H$, we find 
\begin{equation}
N_r(R,H) = \frac{3}{4\pi R^3} \left(1  - \frac{3 R^2 H^2}{10} + \frac{101 R^4 H^4}{1400} +\cdots \right).
\end{equation}

\end{appendix}
%%%
\bibliography{bibliography}{}
\bibliographystyle{utphys}
\end{document}